\documentclass[12pt,english,floatfix,superscriptaddress,aps,preprint,amsmath,amssymb]{revtex4}
\usepackage[latin9]{inputenc}
\usepackage{amsmath}
\usepackage{amssymb} 
\usepackage{amsbsy}
\usepackage{amsfonts}
\usepackage{amsopn}
\usepackage{amstext}
\usepackage{graphicx}
\usepackage{esint}
\usepackage{hyperref}

\usepackage{amssymb}
\usepackage{amsfonts}
\usepackage{amsmath}
\usepackage{graphicx}
\usepackage[english]{babel}
\usepackage{color}
\usepackage[dvips]{epsfig}
\usepackage[dvips]{graphicx}
\usepackage{float}
\usepackage{units}
\usepackage{textcomp}
\usepackage{babel}

\def \be {\begin{equation}}
\def \ee {\end{equation}}
\def \bea{\begin{eqnarray}}
\def \eea{\end{eqnarray}}
\def \ba {\begin{align}}
\def \ea {\end{align}}

\def \a {\alpha}
\def \b {\beta}

\def \G {\Gamma}
\def \d {\delta}

\def \m {\mu}
\def \n {\nu}

\def \l {\lambda}

\def \s {\sigma}

\def \r {\rho}
\def \o {\omega}
\def \O {\Omega}

\def \t {\tau}

\def \p {\partial}
\def \f {\frac}

\def \nn {\nonumber}

\def \la {\leftarrow}

\def \la {\label}

\begin{document}

\title{Disformal transformation in Newton-Cartan geometry}
\author{Peng Huang}
\email{huangp46@mail.sysu.edu.cn}
\affiliation{School of Astronomy and Space Science, Sun Yat-Sen University, Guangzhou
510275, China}

\author{Fang-Fang Yuan}
\email{ffyuan@nankai.edu.cn}
\affiliation{School of Physics, Nankai University, Tianjin 300071, China}

\vskip 1.5cm
\begin{abstract}

Newton-Cartan geometry has played a central role in recent discussions of non-relativistic holography and condensed matter systems.
Although the conformal transformation in non-relativistic holography can be easily rephrased in Newton-Cartan geometry,
we show that it requires a nontrivial procedure to get the consistent form of anisotropic disformal transformation in this geometry. Furthermore,
as an application of the newly obtained disformal transformation, we use it to induce a new geometry.


\end{abstract}

\maketitle

\section{Introduction}

Newton-Cartan geometry (NCG) was proposed by \'Elie Cartan through a geometrical description for Newtonian gravity in the spirit of general relativity \cite{Cartan:1923zea,Cartan:1924yea}.
The recent renewed interests in NCG are well motivated from its usefulness in the investigations of condensed matter systems and non-relativistic holography. Concretely,
it has been found that, together with the non-relativistic diffeomorphism invariance, NCG provides a natural geometrical background for the effective field theory description of fractional quantum Hall effect \cite{Son:2013rqa}; furthermore, the progress in gauge/gravity duality attaches increasing importance to non-Riemannian geometries, specific to NCG, it acts important role in various realizations of non-AdS holography, see e.g. \cite{Christensen:2013lma, Christensen:2013rfa, Hartong:2014oma,Bergshoeff:2014uea,Hartong:2015wxa}.

Similar to the extension from Riemannian geometry to Weyl geometry, the conformal extension of NCG is also investigated from different perspectives. It is noticed that NCG can be obtained by gauging the Bargmann algebra which is the centrally extended Galilean algebra \cite{Andringa:2010it}. Thus the conformal generalization of NCG can be obtained by performing a similar gauging procedure to the Schr\"odinger algebra, the conformal extension of Bargmann algebra \cite{Bergshoeff:2014uea,Hartong:2015zia}. From the perspective of the localization of spacetime symmetries \cite{Blagojevic:2002du,Banerjee:2014pya}, this conformal extension of NCG can also be obtained by localizing the global Galilean and scale symmetries \cite{Mitra:2015twa}.

On the other hand, as a generalization of the conventional conformal transformation, the disformal transformation \cite{Bekenstein:1992pj} also attracts a lot of attention in these years.
It acts important role in Horndeski theories and has led to many important results in modified gravity and cosmology, see \cite{{Horndeski:1974wa},{Deffayet:2011gz},{Gleyzes:2014dya},{Gao:2014soa},{Gao:2014fra},{Bettoni:2013diz},{Zumalacarregui:2012us},{Zumalacarregui:2013pma},{vandeBruck:2015ida},{Koivisto:2015vda},{Koivisto:2015mwa},{Hagala:2015paa},{Brax:2015hma},{Ip:2015qsa}}. Then, a natural question to ask is what is the consistent form (if it exists) of the disformal transformation in NCG and what interesting results will it lead to. While the conformal transformation can be straightforwardly incorporated in NCG and non-relativistic field theories, to find the analogue of disformal transformation is a nontrivial task. This will be the first topic of the present work and it will be shown that the result naturally includes the usual anisotropic conformal rescaling as a special case. Furthermore, as an exploration of its potential usefulness, we use the newly obtained disformal transformation to induce a new geometry.

This paper is organized as follows.
In Sec.(\ref{sec2}), a brief review of the general torsional NCG recently expatiated in \cite{Hartong:2015zia} is given. We also demonstrate that the conformal extension of NCG can be induced through the anisotropic Weyl rescaling of original metrics. This serves as a complementary perspective to the standard gauging or localization procedure. In Sec.(\ref{sec3}), we propose a method to find the most general anisotropic disformal transformation in NCG. Then, as an application of the newly obtained disformal transformation, we use it to induce the disformal extension of NCG in Sec.(\ref{sec4}). The results consist of the basic disformal connection and semi-metricity conditions. Our conclusion will be given in the last section.

\section{Newton-Cartan geometry and its conformal extension} \la{sec2}

\subsection{Newton-Cartan geometry}

A first taste of NCG can be gained through a geometric reformulation of the Newtonian gravity, see \cite{Misner:1974qy} for a clear exposition.
Recently it has been shown that NCG can be obtained by gauging the centrally extended Galilean algebra, i.e. the Bargmann algebra \cite{Andringa:2010it,Bergshoeff:2014uea,Hartong:2015zia}.
Without gauging fixing, the torsion can also be naturally included in this procedure which acts an important role in non-AdS holography.
For example, the boundary geometry in Lifshitz holography is described by torsional NCG \cite{Hartong:2014oma,Hartong:2015wxa}.

Whether there is torsion or not can lead to very different results. While the dynamical NCG without torsion is related to projectable Ho\v rava-Lifshitz gravity,
the incorporation of twistless torsion corresponds to non-projectable Ho\v rava-Lifshitz gravity \cite{Hartong:2015zia}.
 Since the realization of disformal transformation in NCG and its potential applications are of our most concern,
 we would like to maintain as general as possible.
In other words, no gauging fixing on the NCG will be considered and a general torsion (with no twistless condition) is assumed.
For our purpose, it is convenient to make use of the formalism of dynamical NCG developed in \cite{Hartong:2015zia}.
In the following we list some relevant results which will be the starting point of the present work .

One needs two degenerate metrics, $\t_\m$ and $h^{\m\n}$, to define a Newton-Cartan structure on an orientable manifold. Roughly
speaking, $\t_\m$ is a nowhere-vanishing (generally not closed) one-form and defines a local time direction, and $h^{\m\n}$ is a symmetric tensor which gives an inverse metric on the spatial slices.
Both of these two metrics are invariant under internal transformations which include Milne boosts and spatial rotations. Their inverses need to be carefully defined to ensure the invariance of affine connection under these internal transformations.

After all the subtle points have been settled down, one finds that, together with the inverses of $\t_\m$ and $h^{\m\n}$ denoting as $\hat v^\m$ and $H_{\m\n}$ respectively, the most general form of NCG is defined as follows \cite{Hartong:2015zia}:
\be
\la{2.1}
\hat v^\m\t_\m=-1,\quad \hat v^{\m}H_{\m\n}=(2-\a)\tilde \Phi \t_{\n},\quad \t_\m h^{\m\n}=0, \quad h^{\m\n}H_{\m\r}=\d^{\n}_{\r}+\hat v^{\n}\t_{\r},
\ee
with $\a$ an arbitrary constant. The affine connection is defined as
\be
\la{2.2}
\G^{\l}_{\m\n}=-\hat v^{\l}\p_{\m}\t_{\n}+\f{1}{2}h^{\l\s}(\p_{\m}H_{\s\n}+\p_{\n}H_{\s\m}-\p_{\s}H_{\m\n}),
\ee
to obey the metricity postulates
\be
\la{2.3}
\nabla_\m \t_\n=0,\quad \nabla_\m h^{\n\r}=0.
\ee
Here $H_{\m\n}$ is given by
\be
\la{2.4}
H_{\m\n}=\hat h_{\m\n}+(\a-2)\t_\m\t_\n\tilde \Phi.
\ee
Another form of general affine connection can be found in  \cite{{Bekaert:2014bwa}, {Geracie:2015xfa}, {Geracie:2015dea}}.

An intriguing and important point is that despite the difference in the explicit forms of $H_{\m\n}$ and $\hat h_{\m\n}$,
they actually lead to the same results in
constructing the effective actions for Ho\v rava-Lifshitz gravity \cite{Hartong:2015zia}.
Based on this observation, we will choose $\a=2$ from now on. For the sake of convenience, we close this subsection with a rewriting of (\ref{2.1}) and (\ref{2.2}) when this condition is imposed:
\be
\la{2.5}
\hat v^\m\t_\m=-1,\quad \hat v^{\m}\hat h_{\m\n}=0,\quad \t_\m h^{\m\n}=0, \quad h^{\m\n}\hat h_{\m\r}=\d^{\n}_{\r}+\hat v^{\n}\t_{\r},
\ee
\be
\la{2.6}
\G^{\l}_{\m\n}=-\hat v^{\l}\p_{\m}\t_{\n}+\f{1}{2}h^{\l\s}(\p_{\m}\hat h_{\s\n}+\p_{\n}\hat h_{\m\s}-\p_{\s}\hat h_{\m\n}).
\ee

\subsection{Conformal extension of Newton-Cartan geometry} \la{ccc}

In principle, via conformal rescaling (which reduces to Weyl rescaling in the isotropic case), the conformal NCG could be induced from the original geometry.
Our main task in this subsection is to demonstrate this point in detail.

In relativistic systems, 
the space and time have the same footing, and thus the conformal rescaling must be isotropic. Nevertheless, in non-relativistic systems, there is a preferred notion of time. 
So an anisotropic rescaling can be introduced which has the form
\be
\la{2.7}
\t_\m \rightarrow \tilde \t_\m=e^{z\O}\t_\m, \qquad h^{\m\n} \rightarrow \tilde h^{\m\n}=e^{-2\O}h^{\m\n}.
\ee
To preserve the defining relations of NCG in (\ref{2.5}),
the transformation rules for the inverses of $\t_\m$ and $h^{\m\n}$ are constrained as
\be
\la{2.77}
\hat v^\m\rightarrow\tilde v^\m=e^{-z\O}\hat v^\m, \quad
\hat h_{\m\n} \rightarrow  \tilde h_{\m\n}=e^{2\O}\hat h_{\m\n}.
\ee
With these relations at hand, one can perform the following calculations:
\bea
\la{2.8}
\nabla_\m\t_\n=0&=&-ze^{-z\O}\p_\m\O\cdot \tilde \t_\n+e^{-z\O}\nabla_\m\tilde \t_\n , \nn \\
\Rightarrow\nabla_\m \tilde \t_\n&=&z\p_\m\O\cdot \tilde \t_\n; \\
\la{2.9}
\nabla_\m h^{\a\b}=0&=&2e^{2\O}\p_\m\O\cdot \tilde h^{\a\b}+e^{2\O}\nabla_\m\tilde h^{\a\b} ,  \nn \\
\Rightarrow \nabla_\m\tilde h^{\a\b}&=&-2\p_\m\O\cdot\tilde h^{\a\b}.
\eea
The original connection given in (\ref{2.6}) can also be rewritten with new variables as
\bea
\la{2.10}
\G^{\l}_{\m\n}&=&-\tilde v^{\l}(\p_{\m}-z\p_\m\O)\tilde \t_{\n}\nn \\
&&+\f{1}{2}\tilde h^{\l\s}\big[(\p_{\m}-2\p_\m\O)\tilde h_{\s\n}+(\p_{\n}-2\p_\m\O)\tilde h_{\m\s}-(\p_{\s}-2\p_\m\O)\tilde h_{\m\n} \big].
\eea
It is easy to indicate that the expressions in (\ref{2.8})-(\ref{2.10}) are invariant under the transformation $\p_\m\O\rightarrow \p_\m\O+\p_\m\o$.

One can also take a step further and
promote the exact one-form (component form) $\p_\m\O$ into a general one-form $b_\m$.
The resulting equations are still invariant under a similar transformation $b_\m\rightarrow\p_\m\o$.
To exhibit this,
we relabel $\tilde \t_\m$, $\tilde v^\m$, $\tilde h_{\m\n}$ and $\tilde h^{\m\n}$ as $\t_\m$, $\hat v^\m$, $\hat h_{\m\n}$ and $h^{\m\n}$, respectively. One then notices that these quantities satisfy the following conditions:
\be
\la{2.11}
\hat v^\m\t_\m=-1,\quad  \hat v^{\m} \hat h_{\m\n}=0,\quad \t_\m h^{\m\n}=0, \quad h^{\m\n}\hat h_{\m\r}=\d^{\n}_{\r}+ \hat v^{\n}\t_{\r}.
\ee
Furthermore, the semi-metricity conditions (the original metricity postulates are violated) can be rewritten as
\be
\la{2.12}
^{W}\nabla_\m  \t_\n=zb_\m \t_\n, \quad ^{W}\nabla_\m h^{\a\b}=-2b_\m h^{\a\b},
\ee
with the connection given by
\bea
\la{2.13}
^{W}\G^{\l}_{\m\n}&=&-\hat v^{\l}(\p_{\m}-zb_\m) \t_{\n}\nn \\
&&+\f{1}{2} h^{\l\s}\big[ (\p_{\m}-2b_\m) \hat h_{\s\n}+(\p_{\n}-2b_\m) \hat h_{\m\s}-(\p_{\s}-2b_\m) \hat h_{\m\n} \big].
\eea

It is obvious that (\ref{2.11}) assumes the form of the defining relations (\ref{2.5}) for metrics in NCG.
Our notations are thus consistent with the fact that, before and after the conformal rescaling,
these foundamental relations should be maintained.
This also highlights the point that
what one obtained through the procedure (\ref{2.7})-(\ref{2.10}) is a new connection, \textit{not} new metrics.
From the view of basic geometric structures,
the conformal rescaling of metrics for the original NCG may be deemed as just an intermediary step.

We remark that the conformal extension of NCG, whose metric compatibility conditions and connection are given by (\ref{2.11}), (\ref{2.12}) and (\ref{2.13}) respectively, can be looked as a non-relativistic version of the Weyl geometry
(hence the superscripts "W"s in the above equations).
Our results are also consistent with those obtained through the gauging procedure \cite{Bergshoeff:2014uea,Hartong:2015zia}.
Furthermore, when $b_\m$ (which is just the dilatation parameter in \cite{Bergshoeff:2014uea,Hartong:2015zia}) can be expressed as the derivative of a scalar field $\p_\m b$, the corresponding geometry defined by (\ref{2.11})-(\ref{2.13}) is just the Weyl integrable extension of NCG, whose relativistic cousin is the well-known Weyl integrable geometry.

To sum up,
the defining relations (\ref{2.11})-(\ref{2.13}) can be obtained through the
rewriting and rearrangement of the original variables and expressions.
It is just in this sense that we say the conformal extension of NCG can be \textit{induced} from NCG by conformally rescaling the metrics.
The potential advantage of this perspective is that one can use this approach to induce a new geometry
even when the symmetries and algebraic structures of the underlying geometry are not well understood yet.
This method has recently been utilized to induce a new geometry from Riemannian geometry through the disformal transformation
\cite{Yuan:2015tta}.

\section{Anisotropic disformal transformation in Newton-Cartan geometry}   \la{sec3}
The disformal transformation is a generalization of the conventional conformal transformation of metrics which is defined as \cite{Bekenstein:1992pj}
\be
\la{3.1}
\bar g_{\m\n}=A(\phi, X)g_{\m\n}+B(\phi, X)\phi_\m\phi_\n ,
\ee
where the disformal functions $A$ and $B$ depend on the real scalar field $\phi$ and its kinetic term $X=\f{1}{2}g^{\m\n}\phi_\m\phi_\n$ with $\phi_{\m} \equiv \p_\m\phi$. 

By inspection of (\ref{2.7}), (\ref{2.77}) and (\ref{3.1}), a direct generalization 
of anisotropic disformal transformation for NCG would be as follows:
\bea
\la{3.2}
\t_{\m}\rightarrow\bar \t_{\m}&=&A^z\t_{\m}+B\phi_{\m}, \qquad h^{\m\n}\rightarrow\bar h^{\m\n}=A^{-2}h^{\m\n}+C\phi^{\m}\phi^{\n},\nn \\
\hat v^{\m}\rightarrow\bar v^{\m}&=&A^{-z}\hat v^{\m}+E\phi^{\m},\qquad\hat h_{\m\n}\rightarrow \bar h_{\m\n}=A^2\hat h_{\m\n}+D\phi_{\m}\phi_{\n} ,
\eea
where $A$, $B$, $C$, $D$ and $E$ are all functions of the real scalar field $\phi$ and its kinetic term $X$.
It needs to be stressed that, in contrast with $\phi_{\m} \equiv \p_{\m}\phi$, the meaning of $\phi^{\m}$ is not clear at this moment since the form of the metric used to raise the index of $\phi_{\m}$ has not been clarified yet. However, such ambiguity does not obstruct us to denote $X=\f{1}{2}\phi^\m\phi_\m$.
On the other hand, we arrive at the following constraint equations to ensure the defining relations given in (\ref{2.5}):
\bea
\la{3.3}
A^zE\phi^{\m}\t_{\m}+A^{-z}B\hat v^{\m}\phi_{\m}+2BEX&=&0, \\
\la{3.4}
A^zC\t_{\m}\phi^{\m}\phi^{\n}+A^{-2}B\phi_{\m}h^{\m\n}+2BCX\phi^{\n}&=&0,  \\
\la{3.5}
A^{-2}Dh^{\m\n}\phi_{\m}\phi_{\r}+A^2C\phi^{\m}\phi^{\n}\hat h_{\m\r}+2DCX\phi^{\n}\phi_{\r}&&\nn \\
-A^{-z}B\hat v^{\n}\phi_{\r}-A^zE\phi^{\n}\t_{\r}-BE\phi^{\n}\phi_{\r}&=&0,\\
\la{3.6}
A^{-z}D\hat v^{\m}\phi_{\m}\phi_{\n}+A^2E\phi^{\m}\hat h_{\m\n}+2DEX\phi_{\n}&=&0.
\eea

Depending on different definitions of $\phi^\m$, one may encounter the following cases:
\begin{itemize}
\item If $\phi^{\m}=h^{\m\n}\phi_{\m}$ (which is not totally unmotivated although quite peculiar),
then we have $\t_{\m}\phi^{\m}=0$ and (\ref{3.3})-(\ref{3.6}) lead to
    \be
    \label{3.7}
    E=-\f{\phi_{\m}\hat v^{\m}}{2A^zX}, \quad C=-\f{1}{2A^2X},
    \ee
    \be
    \label{3.8}
    (-\f{1}{2X}+\f{B\phi_{\m}\hat v^{\m}}{2A^zX})\phi^{\m}=\f{B}{A^z}\hat v^{\n} ,
    \ee
    \be
    \label{3.9}
    -\phi_\n=\phi_\a\hat v^\a\cdot \t_\n.
    \ee
Obviously since $\phi^\m$ is defined to be independent of $\hat v^\m$, both sides of (\ref{3.8}) must be equal to zero for its validity.
However, this would require $B=0$ and $X=\infty$.
Furthermore, (\ref{3.9}) implies that $\phi^\m$ is identical to zero which is also unacceptable.
\item A more natural choice is $\phi^{\m}=g^{\m\n}\phi_{\n}=(-\hat v^{\m}\hat v^{\n}+h^{\m\n})\phi_{\n}$ (with its inverse defined through $g_{\m\n}=-\t_\m\t_\n+\hat h_{\m\n}$ \cite{Hartong:2015zia,Jensen:2014aia}).
In this case,
generally only when $B=C=D=E=0$ can (\ref{3.3})-(\ref{3.6}) be satisfied,
which just corresponds to the conventional anisotropic conformal transformation. 
If we consider the following novel possibility (with $F$ a function)
    \be
    \la{3.10}
    \phi_{\m} = F \t_{\m}, \quad \phi^{\m} = F \hat v^{\m} ,
    \ee
    then one is left with a particular form of
    disformal transformation 
    (after the constraint equations for coefficients being used)
    \bea
    \la{3.11}
\bar \t_{\m}&=&(A^z+BF)\t_{\m},\quad \bar h^{\m\n}=A^{-2}h^{\m\n},\nn \\
\bar v^{\m}&=&(A^z+BF)^{-1}\hat v^{\m},\quad\bar h_{\m\n}=A^2\hat h_{\m\n}.
\eea
\end{itemize}

One can absorb $F$ into $B$ by redefining for convenience. Notice that $A^z+B$ generally cannot be rewritten as $A^{z'}$ with $z'$ a constant as the new dynamical exponent, thus, (\ref{3.11}) is the desired form of disformal transformation with the assumption of (\ref{3.2}). Furthermore, one can argue that the conventional anisotropic conformal rescaling can be enlarged to general cases that the conformal factors for temporal and spatial components can not be related through the dynamical component $z$ \cite{Mitra:2015twa}. This is in fact effectively equivalent to the case given in (\ref{3.11}),  which is naturally required in seeking a general and consistent form of the disformal transformation in NCG.

Despite such result is consistent from beginning to end, the disturbing aspect of (\ref{3.11}) is that the metric on spatial slices is not involved. 
Apparently with this method we could not find
the most general form of disformal transformation. Thus, we want to tackle this problem from another perspective.
Recall that there is an absolute distinction between space and time in NCG while (\ref{3.2}) does not reflect this character
(especially for $\phi_\m$ and $\phi^\m$).               
Given a nondegenerate metric $g_{\m\n}$, one can introduce the projectors onto the temporal and spatial directions respectively as
\cite{Jensen:2014aia,Hidaka:2014fra}
\be
\la{3.12}
\text{temporal projector: }-\hat v^\m \t_\n,
\quad
\text{spatial projector: }h^\m_\n=h^{\r\m}\hat h_{\r\n}=\d^\m_\n+\hat v^\m\t_\n.
\ee
Correspondingly, $\phi_{\m}$ and $\phi^\m$ can be divided into temporal and spatial parts as
\be
\la{3.13}
\phi_\m=-2Y\t_\m+(\phi_\m+2Y\t_\m),\quad \phi^\m=-2Y\hat v^\m+(\phi^\m+2Y\hat v^\m) ,
\ee
where $Y=\f{1}{2}\hat v^\m\phi_\m=\f{1}{2}\t_\m\phi^\m$. The disformal transformation then takes the following form
\bea
\la{3.14}
\t_{\m}\rightarrow\bar \t_{\m}&=&A^z\t_{\m}+B(-2Y\t_\m),\nn \\
h^{\m\n}\rightarrow\bar h^{\m\n}&=&A^{-2}h^{\m\n}+C(\phi^\m+2Y\hat v^\m)(\phi^\n+2Y\hat v^\n),\nn \\
\hat v^{\m}\rightarrow\bar v^{\m}&=&A^{-z}\hat v^{\m}+E(-2Y\hat v^{\m}), \nn \\
\quad \hat h_{\m\n}\rightarrow\bar h_{\m\n}&=&A^2\hat h_{\m\n}+D(\phi_\m+2Y\t_\m)(\phi_\n+2Y\t_\n).
\eea
Through a similar procedure as above, 
one gets two constraint equations
\bea
\la{3.15}
(A^z-2BY)(A^{-z}-2EY)&=&1,\nn \\
A^{-2}D+A^2C+DC(2X+4Y^2)&=&0 .
\eea
Together with (\ref{3.14}), the final form of the disformal transformation in NCG is obtained as
\bea
\la{3.16}
\bar \t_\m&=&(A^z-2BY)\t_{\m},\nn \\
\bar h^{\m\n}&=&A^{-2}h^{\m\n}+C\Phi^\m\Phi^\n,\nn \\
\bar v^\m&=&(A^z-2BY)^{-1}\hat v^{\m},\nn \\
\bar h_{\m\n}&=&A^2\hat h_{\m\n}-\f{A^4C}{1+2A^2CZ}\Phi_\m\Phi_\n .
\eea
We have defined $\Phi_\m  \equiv  \phi_\m+2Y\t_\m$, $\Phi^\m=\Phi_\n g^{\m\n}= \phi^\m+2Y\hat v^\m$ and $Z=X+2Y^2=\f{1}{2}\Phi^\m\Phi_\m$.

Before closing this section, several comments are in order:
\begin{itemize}
\item The first two equations in (\ref{3.16}) suffice to give the definition of the related disformal transformation since their inverses can be derived from them and thus not independent.

\item The previous result in (\ref{3.11}) simply corresponds to the special case when $Y=-\f{1}{2} F$, which also means $\phi_\m$ has only temporal components.

\item With respect to the disformal transformation (\ref{3.1}) in relativistic gravity theories, it has been shown that $A(\phi,X)$ must be positive definite, and $B(\phi,X)$ should satisfy some constraints to ensure a healthy definition \cite{Bekenstein:1992pj,Bettoni:2013diz,Zumalacarregui:2012us}.
The similar situation is also expected to happen in non-relativistic cases.
Nevertheless, this is not of the present concern and is left for future study.
\end{itemize}

\section{Disformal extension of Newton-Carton geometry}    \la{sec4}

As an application of the newly obtained disformal transformation (\ref{3.16}), we would like to use it to induce a new geometry which is the disformal extension of NCG.
Noticing that the relativistic disformal transformation in the form of $\bar g_{\a\b}=A(\phi)g_{\a\b}+B(\phi)\phi_{\a}\phi_{\b}$ is of particularly interest in recent literature \cite{Zumalacarregui:2012us,Bettoni:2013diz,Zumalacarregui:2013pma,vandeBruck:2015ida,Koivisto:2015vda,Koivisto:2015mwa,Hagala:2015paa,Brax:2015hma,Ip:2015qsa},
we will restrict our attention to the special case where the disformal parameters $A$, $B$ and $C$ in (\ref{3.16}) are only functions of the scalar field $\phi$ throughout this section.

The key step in inducing a new geometry is to rewrite the original geometrical variables with new variables obtained through disformal transformation.
To implement this, one firstly notices that $\bar Y=\f{1}{2}\bar v^\m\phi_\m=(A^z-2BY)^{-1}Y$ which leads to
\be
\la{4.2}
Y=\f{A^z\bar Y}{1+2B\bar Y}.
\ee
With this relation, it is easy to get
\bea
\la{4.3}
\t_\m&=&A^{-z}(1+2B\bar Y)\bar \t_\m, \\
\hat v^\m&=&A^z(1+2B\bar Y)^{-1}\bar v^\m,\\
\phi_\m&=&-2Y\t_\m+(\phi_\m+2Y\t_\m)\nn\\
&=&-2\bar Y\bar \t_\m+(\phi_\m+2\bar Y\bar \t_\m).
\eea
A subtle point is that with the definition $\bar h^{\m\n}=A^{-2}h^{\m\n}+C\Phi^\m\Phi^\n$ at hand,
rewriting $h^{\m\n}$ with $\bar h^{\m\n}$ and $\Phi^\m$ will inevitably lead to an infinite iteration due to the hidden $h^{\m\n}$ in $\Phi^\m$.
The way to circumvent this problem is to notice that the disformal transformation of $h^{\m\n}$ and $\hat h_{\m\n}$ can be defined in an equivalent form as
\bea
\la{4.3}
\bar h^{\m\n}&=&A^{-2}h^{\m\n}-\f{D}{A^4+2A^2ZD}\Phi^\m\Phi^\n,\nn \\
\bar h_{\m\n}&=&A^2\hat h_{\m\n}+D\Phi_\m\Phi_\n,
\eea
where we have introduced new coefficients $D(\phi)$,
Then one immediately gets
\be
\la{4.4}
\hat h_{\m\n}=A^{-2}\bar h_{\m\n}-\f{D}{A^2}\Phi_\m\Phi_\n.
\ee
The inverse of $\hat h_{\m\n}$ is supposed to have the form $h^{\m\n}=A^2\bar h^{\m\n}+G\bar\Phi^\m\bar\Phi^\n$,
where $\bar\Phi^\m\equiv \Phi_\n \bar g^{\m\n}=\Phi_\n \bar h^{\m\n}$ and $G$ is an unknown function to be determined.
It is straightforward to find that
\be
\la{4.5}
h^{\m\n}=A^2\bar h^{\m\n}+\f{A^2D}{1-2D\bar Z}\bar\Phi^\m\bar\Phi^\n
\ee
with $\bar Z=\f{1}{2}\bar \Phi^\m\Phi_\n$.

Based on the above results, through some straightforward calculations,
we find that the original connection (\ref{2.6}) can be expressed in terms of the new variables as
\bea
\la{4.6}
\G^\l_{\m\n}&=&-\bar v^\l  \bigg[  \p_\m-\p_\m   \Big(   \ln \f{A^z}{1+2B\bar Y}  \Big)   \bigg]  \bar \t_\n\nn \\
&&+\f{1}{2}  \Big(\bar h^{\l\s}+\f{D}{1-2D\bar Z}\bar \Phi^\l\bar \Phi^\s \Big) \Big[(\p_\m-\p_\m\ln A^2)(\bar h_{\s\n}-D\Phi_\s\Phi_\n)\nn \\
&&+(\p_\n-\p_\n\ln A^2)(\bar h_{\m\s}-D\Phi_\m\Phi_\s)-(\p_\s-\p_\s\ln A^2)(\bar h_{\m\n}-D\Phi_\m\Phi_\n)\Big] .
\eea

Furthermore, the semi-metricity conditions for the new geometry can also be induced.
For $\bar \t_\m$, the result is
\bea
\la{4.8}
\nabla_\m\t_\n=0&=&\nabla_\m\bigg(\f{1+2B_2\bar Y}{A^z}\bar \t_\n\bigg),   \nn \\
\Rightarrow\nabla_\m \bar \t_\n&=&\p_\m \bigg(\ln\f{A^z}{1+2B\bar Y}\bigg)\cdot \bar \t_\n .
\eea
The semi-metricity condition for $\bar h^{\a\b}$ on the other hand takes a more complicated form. To induce it, the first step is to rearrange the original metric compatibility condition as
\bea
\la{4.9}
\nabla_\m h^{\a\b}=0&=&\nabla_\m  \bigg[   \Big(A^2\bar h^{\a\b}+\f{A^2D}{1-2D\bar Z}  \Big)\bar \Phi^\a\bar\Phi^\b\bigg] , \nn \\
\Rightarrow \nabla_\m  \Big(\bar h^{\a\b}&+&\f{D}{1-2D\bar Z}\bar \Phi^\a\bar\Phi^\b\Big)=-\p_\m \ln A^2\cdot\Big(\bar h^{\a\b}+\f{D}{1-2D\bar Z}\bar \Phi^\a\bar\Phi^\b\Big).
\eea

As been clearly demonstrated in the case of conformal extension of NCG,
while we have induced a new connection by implementing the disformal transformation of the original metrics as an intermediary step,
the metrics themselves in fact do not change.
Thus for clarity it is more proper to rewrite $\bar \t_\m$, $\bar v^\m$, $\bar h_{\m\n}$ and $\bar h^{\m\n}$ as
$\t_\m$, $\hat v^\m$, $\hat h_{\m\n}$ and $h^{\m\n}$, respectively.
Then it can been seen that the new induced geometry has two degenerate metrics, $\t_\m$ and $ h^{\m\n}$, which satisfy (as in (\ref{2.5}))
\be
\la{4.11}
\hat v^\m\t_\m=-1,\quad \hat v^{\m} \hat h_{\m\n}=0,\quad \t_\m h^{\m\n}=0, \quad h^{\m\n}\hat h_{\m\r}=\d^{\n}_{\r}+ \hat v^{\n}\t_{\r} .
\ee
The corresponding connection and semi-metricity conditions are given as follows:
\begin{itemize}
\item Disformal connection:
\bea
\la{4.12}
^D\G^\l_{\m\n}&=&- \hat v^\l  \bigg[  \p_\m-\p_\m   \Big(   \ln \f{A^z}{1+2B Y}  \Big)   \bigg]   \t_\n\nn \\
&&+\f{1}{2}  \Big( h^{\l\s}+\f{D}{1-2D Z} \Phi^\l \Phi^\s \Big) \Big[(\p_\m-\p_\m\ln A^2)(\hat h_{\s\n}-D\Phi_\s\Phi_\n)\nn \\
&&+(\p_\n-\p_\n\ln A^2)(\hat h_{\m\s}-D\Phi_\m\Phi_\s)-(\p_\s-\p_\s\ln A^2)( \hat h_{\m\n}-D\Phi_\m\Phi_\n)\Big] .
\eea
Here $\p_\m-\p_\m   \big(\ln \f{A^z}{1+2B Y} \big)$ can be interpreted as the Weyl covariant derivative along the time direction,
and $\p_\m-\p_\m\ln A^2$ is the Weyl covariant derivative on the spatial slices.
For the special case with $B=D=0$, (\ref{4.12}) reduces to (\ref{2.13}) which is the result obtained in the conformal extension of NCG.

\item Semi-metricity conditions:
\bea
\la{4.13}
^D\nabla_\m  \t_\n&=&\p_\m \bigg(\ln\f{A^z}{1+2B Y}\bigg)\cdot  \t_\n, \\
\la{4.14}
^D\nabla_\m  \Big( h^{\a\b}&+&\f{D}{1-2D Z} \Phi^\a\Phi^\b\Big)=-\p_\m \ln A^2\cdot\Big( h^{\a\b}+\f{D}{1-2D Z} \Phi^\a\Phi^\b\Big).
\eea
These two semi-metricity conditions with the disformal connection (\ref{4.12}) define the disformal extension of the NCG.  Once again, the conformal extension of NCG (\ref{2.12}) is naturally included here as a special case. Furthermore, geometry defined by (\ref{4.12})-(\ref{4.14}) is the nonrelativistic and disformally extended version of the conventional Weyl integrable geometry.
\end{itemize}


\section{Conclusion and Discussion}

In this work, we have found the general anisotropic disformal transformation (\ref{3.16}) in NCG. It is shown that a naive assumption of the form of the disformal transformation can only lead to a very special form from the consideration of consistency, see (\ref{3.11}). To obtain its most general form, one needs to project the vector $\phi_\m$ in the relativistic disformal transformation (\ref{3.1}) into temporal and spatial parts, and then imposing the constraints from the defining relations in NCG.

As an application, this newly obtained disformal transformation has been used to induce a new geometry whose disformal connection and semi-metricity conditions are given by (\ref{4.12}) and (\ref{4.13}, \ref{4.14}) respectively. The key step to do this is to rewrite the original geometrical variables with new variables obtained through disformal transformation. Consequent subtle point is the infinite iteration in rewriting $h^{\m\n}$ with $ h^{\m\n}$ and $\Phi^\m$ due to the hidden $h^{\m\n}$ in $\Phi^\m$. This problem is circumvented by using another equivalent (but convenient for rewriting) form of the disformal transformation, see (\ref{4.3}).

Take a more look at the semi-metricity conditions for the induced geometry, one can find that (\ref{4.13}, \ref{4.14}) could be rewritten in a more illuminating form
\bea
\la{4.15}
\bigg(\ ^D\nabla_\m-\p_\m \big(\ln\f{A^z}{1+2B Y}\big)\bigg)\t_\n&=&0, \\
\la{4.16}
\big(\ ^D\nabla_\m+\p_\m \ln A^2\big)  \Big( h^{\a\b}+\f{D}{1-2D Z} \Phi^\a\Phi^\b\Big)&=&0.
\eea
In this form, all the derivatives on the L.H.S are adjusted into Weyl covariant type, which is to say that these two equations in fact serve as the non-relativistic and disformal generalization of the metricity conditions for Weyl gauge theory \cite{Blagojevic:2002du}.

For further investigation, it would be worth to explore the implications of disformally coupled fields in Newton-Cartan gravity and its holographic applications.
Although the relevant Lie algebra structure has not been fully understood yet,
it should be possible to obtain the disformal NCG via a gauging procedure.
Finally, it may be interesting to investigate some explicit metric solutions in NCG-based holography.


\begin{thebibliography}{0}

\bibitem{Cartan:1923zea}
  E.~Cartan,
  ``Sur les vari\'et\'es \`a connexion affine et la th\'eorie de la relativit\'e g\'en\'eralis\'ee. (premi\'ere partie),''
  Annales Sci.\ Ecole Norm.\ Sup.\  {\bf 40}, 325 (1923).

\bibitem{Cartan:1924yea}
  E.~Cartan,
  ``Sur les vari\'et\'es \`a connexion affine et la t\'eorie de la relativit\'e g\'en\'eralis\'ee. (premi\'ere partie) (Suite).,''
  Annales Sci.\ Ecole Norm.\ Sup.\  {\bf 41}, 1 (1924).

\bibitem{Son:2013rqa}
  D.~T.~Son,
  ``Newton-Cartan Geometry and the Quantum Hall Effect,''
  arXiv:1306.0638 [cond-mat.mes-hall].

\bibitem{Christensen:2013lma}
  M.~H.~Christensen, J.~Hartong, N.~A.~Obers and B.~Rollier,
  ``Torsional Newton-Cartan Geometry and Lifshitz Holography,''
  Phys.\ Rev.\ D {\bf 89}, 061901 (2014)
  [arXiv:1311.4794 [hep-th]].

\bibitem{Christensen:2013rfa}
  M.~H.~Christensen, J.~Hartong, N.~A.~Obers and B.~Rollier,
  ``Boundary Stress-Energy Tensor and Newton-Cartan Geometry in Lifshitz Holography,''
  JHEP {\bf 1401}, 057 (2014)
  [arXiv:1311.6471 [hep-th]].

\bibitem{Hartong:2014oma}
  J.~Hartong, E.~Kiritsis and N.~A.~Obers,
  ``Lifshitz space-times for Schr\"odinger holography,''
  Phys.\ Lett.\ B {\bf 746}, 318 (2015)
  [arXiv:1409.1519 [hep-th]].

\bibitem{Bergshoeff:2014uea}
  E.~A.~Bergshoeff, J.~Hartong and J.~Rosseel,
  ``Torsional Newton-Cartan geometry and the Schr\"odinger algebra,''
  Class.\ Quant.\ Grav.\  {\bf 32}, no. 13, 135017 (2015)
  [arXiv:1409.5555 [hep-th]].

\bibitem{Hartong:2015wxa}
  J.~Hartong, E.~Kiritsis and N.~A.~Obers,
  ``Field Theory on Newton-Cartan Backgrounds and Symmetries of the Lifshitz Vacuum,''
  JHEP {\bf 1508}, 006 (2015)
  [arXiv:1502.00228 [hep-th]].

\bibitem{Andringa:2010it}
  R.~Andringa, E.~Bergshoeff, S.~Panda and M.~de Roo,
  ``Newtonian Gravity and the Bargmann Algebra,''
  Class.\ Quant.\ Grav.\  {\bf 28}, 105011 (2011)
  [arXiv:1011.1145 [hep-th]].

\bibitem{Hartong:2015zia}
  J.~Hartong and N.~A.~Obers,
  ``Ho\v rava-Lifshitz gravity from dynamical Newton-Cartan geometry,''
  JHEP {\bf 1507}, 155 (2015)
  [arXiv:1504.07461 [hep-th]].

\bibitem{Blagojevic:2002du}
  M.~Blagojevic,
  ``Gravitation and gauge symmetries,''
  Bristol, UK: IOP (2002) 522 p.

\bibitem{Banerjee:2014pya}
  R.~Banerjee, A.~Mitra and P.~Mukherjee,
  ``A new formulation of non-relativistic diffeomorphism invariance,''
  Phys.\ Lett.\ B {\bf 737}, 369 (2014)
  [arXiv:1404.4491 [gr-qc]].

\bibitem{Mitra:2015twa}
  A.~Mitra,
  ``Weyl rescaled Newton-Cartan geometry from the localization of Galilean and scale symmetries,''
  arXiv:1508.03207 [hep-th].

\bibitem{Bekenstein:1992pj}
  J.~D.~Bekenstein,
  ``The Relation between physical and gravitational geometry,''
  Phys.\ Rev.\ D {\bf 48}, 3641 (1993)
  [gr-qc/9211017].

\bibitem{Horndeski:1974wa}
  G.~W.~Horndeski,
  ``Second-order scalar-tensor field equations in a four-dimensional space,''
  Int.\ J.\ Theor.\ Phys.\  {\bf 10}, 363 (1974).

\bibitem{Deffayet:2011gz}
  C.~Deffayet, X.~Gao, D.~A.~Steer and G.~Zahariade,
  ``From k-essence to generalised Galileons,''
  Phys.\ Rev.\ D {\bf 84}, 064039 (2011)
  [arXiv:1103.3260 [hep-th]].

\bibitem{Gleyzes:2014dya}
  J.~Gleyzes, D.~Langlois, F.~Piazza and F.~Vernizzi,
  ``Healthy theories beyond Horndeski,''
  arXiv:1404.6495 [hep-th].

\bibitem{Gao:2014soa}
  X.~Gao,
  ``Unifying framework for scalar-tensor theories of gravity,''
  Phys.\ Rev.\ D {\bf 90}, 081501 (2014)
  [arXiv:1406.0822 [gr-qc]].

\bibitem{Gao:2014fra}
  X.~Gao,
  ``Hamiltonian analysis of spatially covariant gravity,''
  Phys.\ Rev.\ D {\bf 90}, 104033 (2014)
  [arXiv:1409.6708 [gr-qc]].

\bibitem{Zumalacarregui:2012us}
  M.~Zumalacarregui, T.~S.~Koivisto and D.~F.~Mota,
  ``DBI Galileons in the Einstein Frame: Local Gravity and Cosmology,''
  Phys.\ Rev.\ D {\bf 87}, 083010 (2013)
  [arXiv:1210.8016 [astro-ph.CO]].

\bibitem{Bettoni:2013diz}
  D.~Bettoni and S.~Liberati,
  ``Disformal invariance of second order scalar-tensor theories: Framing the Horndeski action,''  Phys.\ Rev.\ D {\bf 88}, 084020 (2013)  [arXiv:1306.6724 [gr-qc]].


\bibitem{Zumalacarregui:2013pma}
  M.~Zumalacarregui and J.~Garcia-Bellido,
  ``Transforming gravity: from derivative couplings to matter to second-order scalar-tensor theories beyond the Horndeski Lagrangian,''  Phys.\ Rev.\ D {\bf 89}, 064046 (2014)  [arXiv:1308.4685 [gr-qc]].

\bibitem{vandeBruck:2015ida}
  C.~van de Bruck and J.~Morrice,
  ``Disformal couplings and the dark sector of the universe,''
  JCAP {\bf 1504}, no. 04, 036 (2015)
  [arXiv:1501.03073 [gr-qc]].


\bibitem{Koivisto:2015vda}
  T.~S.~Koivisto and F.~R.~Urban,
  ``Doubly-boosted vector cosmologies from disformal metrics,''
  Phys.\ Scripta {\bf 90}, no. 9, 095301 (2015)
  [arXiv:1503.01684 [astro-ph.CO]].


\bibitem{Koivisto:2015mwa}
  T.~Koivisto and H.~J.~Nyrhinen,
  ``Stability of disformally coupled accretion disks,''
  arXiv:1503.02063 [gr-qc].


\bibitem{Hagala:2015paa}
  R.~Hagala, C.~Llinares and D.~F.~Mota,
  ``Cosmological simulations with disformally coupled symmetron fields,''
  arXiv:1504.07142 [astro-ph.CO].

\bibitem{Brax:2015hma}
  P.~Brax, C.~Burrage and C.~Englert,
  ``Disformal dark energy at colliders,''
  Phys.\ Rev.\ D {\bf 92}, no. 4, 044036 (2015)
  [arXiv:1506.04057 [hep-ph]].

\bibitem{Ip:2015qsa}
  H.~Y.~Ip, J.~Sakstein and F.~Schmidt,
  ``Solar System Constraints on Disformal Gravity Theories,''
  arXiv:1507.00568 [gr-qc].



\bibitem{Misner:1974qy}
  C.~W.~Misner, K.~S.~Thorne and J.~A.~Wheeler,
  ``Gravitation,''
  San Francisco 1973, 1279p, Chapter 12.

\bibitem{Bekaert:2014bwa}
  X.~Bekaert and K.~Morand,
  ``Connections and dynamical trajectories in generalised Newton-Cartan gravity I. An intrinsic view,''
  arXiv:1412.8212 [hep-th].

\bibitem{Geracie:2015xfa}
  M.~Geracie, K.~Prabhu and M.~M.~Roberts,
  ``Fields and fluids on curved non-relativistic spacetimes,''
  JHEP {\bf 1508}, 042 (2015)
  [arXiv:1503.02680 [hep-th]].

\bibitem{Geracie:2015dea}
  M.~Geracie, K.~Prabhu and M.~M.~Roberts,
  ``Curved non-relativistic spacetimes, Newtonian gravitation and massive matter,''
  arXiv:1503.02682 [hep-th].

\bibitem{Yuan:2015tta}
  F.~-F.~Yuan and P.~Huang,
  ``Induced geometry from disformal transformation,''
  Phys.\ Lett.\ B {\bf 744}, 120 (2015)
  [arXiv:1501.06135 [gr-qc]].

\bibitem{Jensen:2014aia}
  K.~Jensen,
  ``On the coupling of Galilean-invariant field theories to curved spacetime,''
  arXiv:1408.6855 [hep-th].

\bibitem{Hidaka:2014fra}
  Y.~Hidaka, T.~Noumi and G.~Shiu,
  ``Effective field theory for spacetime symmetry breaking,''
  Phys.\ Rev.\ D {\bf 92}, no. 4, 045020 (2015)
  [arXiv:1412.5601 [hep-th]].


\end{thebibliography}
\end{document}